\documentclass[universe,article,submit,pdftex,moreauthors]{Definitions/mdpi} 
\firstpage{1} 
\pubvolume{1}
\issuenum{1}
\articlenumber{0}
\pubyear{2025}
\copyrightyear{2025}
\externaleditor{Eric Schlegel}
\datereceived{18 December 2024} 
\daterevised{22 January 2025} 
\dateaccepted{25 January 2025} 
\datepublished{ } 
\hreflink{https://doi.org/} 

\Title{Constraining the Milky Way’s Dispersion Measure Using FRB and X-ray Data}

\TitleCitation{Constraining the Milky Way’s Dispersion Measure Using FRB and X-ray Data}

\Author{Jiale Wang $^{1}$\orcidA{}, Zheng Zhou $^{1}$\orcidB{}, Xiaochuan Jiang $^{2}$\orcidC{}, Taotao Fang $^{1,*}$\orcidD{}}

\AuthorNames{Jiale Wang, Zheng Zhou, Xiaochuan Jiang and Taotao Fang}

\AuthorCitation{Wang, J.; Zhou, Z.; Jiang, X.; Fang, T.}

\address{%
$^{1}$ \quad Department of Astronomy, Xiamen University, Xiamen 361005, China; wangjiale@stu.xmu.edu.cn (J.W.); zhengz@stu.xmu.edu.cn (Z.Z.)%

$^{2}$ \quad School of Information Engineering, Fujian Business University, Fuzhou 350506, China; jiangxc@xmu.edu.cn}

\corres{Correspondence: \href{mailto:}{fangt@xmu.edu.cn}}

\abstract{The dispersion measures (DMs) of the fast radio bursts (FRBs) are a valuable tool to probe the baryonic content of the intergalactic medium and the circumgalactic medium of the intervening galaxies along the sightlines. However, interpreting the DMs is complicated by the contribution from the hot gas in and around our Milky Way. This study examines the relationship between \(DM_{\text{MW}}\), derived from localized FRBs, and the Galaxy's hot gas, using X-ray absorption and emission data from O \textsc{vii} and O \textsc{viii}. We find evidence for a positive correlation between \(DM_{\text{MW}}\) and O \textsc{vii} absorption, reflecting contributions from both the disk and halo components. This conclusion is supported by two lines of evidence: (1) No correlation between \(DM_{\text{MW}}\) and O \textsc{vii}/O \textsc{viii} emission, which primarily traces dense disk regions; and (2) the comparison with electron density models, where \(DM_{\text{MW}}\) aligns with models that incorporate both disk and halo components but significantly exceeds predictions from pure disk-only models, emphasizing the halo's role. Furthermore, the lack of correlation with O \textsc{viii} absorption suggests that the primary temperature of the Galaxy's hot gas is likely around \(2 \times 10^6\) K or less, as traced by O \textsc{vii} absorption, while gas at higher temperatures (\(\sim 3 - 5 \times 10^6\) K) is present but less abundant. Our findings provide insights into the Milky Way's gas distribution and improve \(DM_{\text{MW}}\) estimates for future cosmological studies.}


\usepackage{lineno}
\renewcommand{\linenumbers}{}

\begin{document}

\section{Introduction} \label{sec:intro}
Fast Radio Bursts (FRBs) are intense, millisecond-duration radio pulses originating from extragalactic sources \citep{lorimer2007}. They often exhibit high dispersion measures (DM), which reflect the integrated electron density along their line of sight. This makes FRBs invaluable tools for probing baryon density across the universe, including addressing the ‘‘missing baryon'' problem, where a significant fraction of baryons at low redshifts are believed to be undetected \citep{Shull2012}. These baryons are thought to exist in the form of diffuse, hot gas within the Intergalactic Medium (IGM) and the Circumgalactic Medium (CGM) of intervening galaxies, which are challenging to observe directly due to their high temperatures, low densities, and the limitations of current telescopes \citep{Gupta_2012,fang2013ApJ...762...20F, nicastro2023ApJ...955L..21N}. By using FRBs to trace the distribution of ionized baryons, we can gain insights into these otherwise elusive components of the universe.

One of the primary challenges in using FRBs for cosmological studies is accurately determining the Milky Way (MW) contribution to the observed DM (\(DM_{\text{MW}}\)). The total observed DM of an FRB can be expressed as \citep{Macquart2020}:

\begin{equation}\label{eq1}
DM_{\text{obs}} = \frac{DM_{\text{Host}}}{1+z} + DM_{\text{IGM}} + DM_{\text{MW,disk}} + DM_{\text{MW,halo}}.
\end{equation}

Here, \(DM_{\text{Host}}\) represents the contribution from the host galaxy, scaled by its redshift (\(z\)), \(DM_{\text{IGM}}\) is the contribution from the intergalactic medium, \(DM_{\text{MW,disk}}\) represents the contribution from the Milky Way disk, and \(DM_{\text{MW,halo}}\) represents the contribution from the Milky Way halo. The sum of \(DM_{\text{MW,disk}}\) and \(DM_{\text{MW,halo}}\) represents the total contribution from the Milky Way, \(DM_{\text{MW}}\). In most previous studies, this total MW contribution has been modeled using a disk component (modeled by NE2001 or YMW16 \citep[see,][]{ne2001.2002astro.ph..7156C, ymw16.2017ApJ...835...29Y}) plus an assumed constant DM\(_{\text{MW,halo}}\) to account for the halo component.

However, this approach can introduce significant uncertainties. The disk and halo components of the Milky Way are highly complex, with the gas distribution being neither uniform nor well-constrained. Simply adding a constant value for the halo contribution overlooks variations in density, temperature, and geometry, leading to potential under- or overestimation of \(DM_{\text{MW}}\). Furthermore, the NE2001 and YMW16 models, widely used for Galactic electron density studies, show limitations in high-latitude regions, produce significant discrepancies in certain directions, and exhibit notable biases when compared to independent pulsar distance measurements, leading to estimation errors in FRB studies \citep{price2021comparison}. As a result, this method can lead to significant inaccuracies in determining the Milky Way's DM contribution, which directly impacts our ability to accurately isolate and study the IGM and CGM components \citep[see,][]{gaensler2008vertical, jennings2018binary}.


These uncertainties have prompted various studies to place constraints on \( DM_{\text{MW,halo}} \) using FRBs. \citet{platts2020data} first established conservative constraints on \( DM_{\text{MW,halo}} \) by analyzing a population of FRBs and adjusting for \( DM_{\text{MW,disk}} \) as computed by the NE2001 model. Their analysis suggested a wide range for \( DM_{\text{MW,halo}} \), between \(\text{-}2\) and \(123 \, \text{pc} \, \text{cm}^{\text{-3}}\), to account for potential uncertainties. Following this, \citet{cook2023frb} refined these estimates using the CHIME/FRB dataset, establishing \( DM_{\text{MW,halo}} \) upper limits between \(52\) and \(111 \, \text{pc} \, \text{cm}^{\text{-3}}\) for Galactic latitudes \(|b| \geq 30^\circ\). More recently, \citet{wei2023investigating} analyzed DM and redshift data from 24 localized FRBs to explore cosmological models. By applying a flat prior for \( DM_{\text{MW,halo}} \) ranging from 5 to 80 \(\text{pc} \, \text{cm}^{\text{-3}}\), they yielded a Hubble constant of \( H_0 = 95.8^{\text{+9.2}}_{\text{-7.8}} \, \text{km s}^{\text{-1}} \, \text{Mpc}^{\text{-1}} \), highlighting the potential of FRBs to contribute to cosmological parameter estimation. In the latest study, \citet{huang2024modeling} employed cosmological hydrodynamical simulations of the Local Universe to construct a DM model encompassing the Milky Way halo, the Local Group (including contributions from the intra-group medium), and the broader Local Universe (up to 120 Mpc). They reported an average value of \( \text{DM}_{\text{MW, halo}} \) of 46 \(\text{pc} \, \text{cm}^{\text{-3}}\) with a standard deviation of 17 \(\text{pc} \, \text{cm}^{\text{-3}}\). These variations in \( DM_{\text{MW,halo}} \) are particularly critical, as the halo component contributes significantly to the overall \( DM_{\text{MW}} \) budget and directly affects our ability to isolate the IGM contribution. Therefore, understanding and constraining \( DM_{\text{MW,halo}} \) more accurately is crucial for reducing uncertainties in the IGM and CGM components, which are key to our understanding of the large-scale structure of the universe.

In this work, we focus on studying the total \( DM_{\text{MW}} \), considering the combined contributions from both the disk and halo components. The Milky Way’s hot gas, with a temperature of a few million degrees, is ideal for producing highly ionized species such as O \textsc{vii} and O \textsc{viii}\citep{smith2007suzaku,Henley2013ApJ...773...92H}. While directly tracing electron density is challenging, these ions serve as effective tracers of the hot, highly ionized gas, as demonstrated by numerous X-ray observations \citep{nicastro2018Natur.558..406N}. By analyzing O \textsc{vii} and O \textsc{viii} emission and absorption data, we aim to establish correlations between these X-ray tracers and the overall \( DM_{\text{MW}} \). This approach will help reduce uncertainties in the Milky Way DM contribution, thereby enhancing the utility of FRBs as cosmological probes. It can also help improve our understanding of the ionized gas distribution within the MW.

The structure of this paper is as follows: In Section \ref{sec:2}, we present a list of localized FRBs and describe the method used to calculate \(DM_{\text{MW}}\). Section \ref{sec:3} focuses on the relationship between FRBs and key Galactic gas tracers, including X-ray O \textsc{vii} and O \textsc{viii} data. In Section \ref{sec:4}, we discuss the observed correlations and compare them with existing MW electron distribution models, and the last section provides a summary.

\section{Data and Methods} \label{sec:2}
\subsection{Localized FRBs Sample}

\begin{figure}[H]
\begin{adjustwidth}{-\extralength}{0cm}
    \centering
    \begin{minipage}{0.75\textwidth} 
        \centering
        \includegraphics[width=\linewidth]{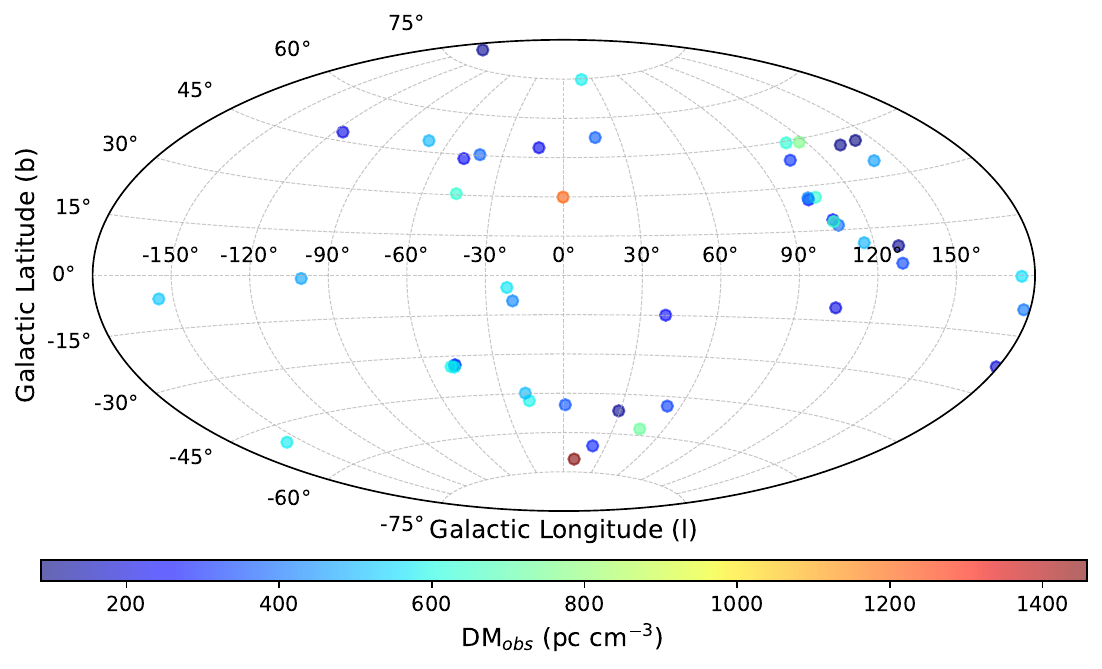}
    \end{minipage}
    \hfill
    \begin{minipage}{0.55\textwidth} 
        \centering
        \includegraphics[width=\linewidth]{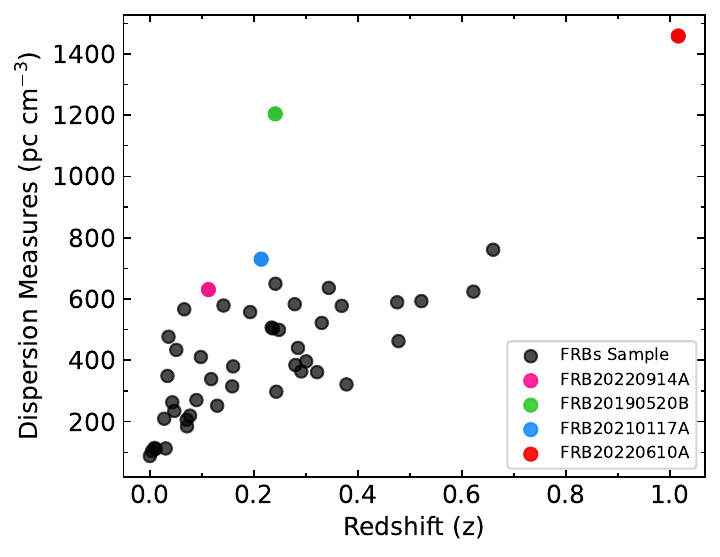}
    \end{minipage}
\end{adjustwidth}
    \caption{Left: Sky distributions in celestial coordinates of 47 localized FRBs analyzed in this work. The color represents the magnitude of their observed DM (DM$_{obs}$). Right: DM as a function of redshift for these FRBs, with four unusually high-DM FRBs highlighted. 
\label{fig:skydis}}
\end{figure}

Our FRB data were obtained from the Blinkverse Database \citep{blinkverse}, which provides a comprehensive and up-to-date collection of FRB observations from various observatories, including FAST, CHIME, GBT, and Arecibo. To identify localized FRBs with known redshifts, we searched for each event in the database using the Astrophysical Data Service (ADS)  \citep{kurtz2000nasa}. This process resulted in a sample of 47 localized FRBs, which are listed in Table \ref{tab:localized frbs} and shown in their sky distribution and as a function of redshift in Figure \ref{fig:skydis}.

\subsection{Derivation of \texorpdfstring{$DM_{\text{MW}}$}{DM\_MW}}\label{DMMW}

Based on the observed \(DM\) values (\(DM_{\text{obs}}\)) of 47 localized FRBs, we derived \(DM_{\text{MW}}\) by subtracting \(DM_{\text{IGM}}\), estimated using average values from the Macquart relation \citep[see,][]{Macquart2020}, and \(DM_{\text{host}}\), assumed to be \(30/(1+z) \, \text{pc cm}^{\text{-3}}\) to account for redshift \(z\), consistent with results from the Illustris-1 simulation, where \citet{mo2023MNRAS.518..539M} reported a median \(DM_{\text{host}}\) value of 31 \(\text{pc cm}^{\text{-3}}\) in their stellar mass model. \(DM_{\text{MW}}\) was derived based on Equation \ref{eq1}. 

To account for possible higher \(DM_{\text{host}}\) values, we also tested assumptions of \(DM_{\text{host}}\) up to 60~\(\text{pc cm}^{\text{-3}}\) and 100~\(\text{pc cm}^{\text{-3}}\). These alternative values were included in the calculations, and while the detailed results are presented as Alternative 1 and 2 in Table \ref{tab:pearson_results}, we emphasize that the choice of \(DM_{\text{host}} = 30~ \text{pc cm}^{\text{-3}}\) does not qualitatively affect our findings.

Notably, this approach yielded unusually high \( DM_{\text{MW}} \) values (exceeding 500 \(\text{pc} \, \text{cm}^{\text{-3}}\)) for FRB 190520B, FRB 210117A, FRB 220610A, and FRB 220914A. Since these FRBs are located at Galactic latitudes above 15 degrees, where the Milky Way’s contribution should be minimal, the observed \( DM_{\text{MW}} \) values far exceed expectations, suggesting significant additional \( DM \) contributions from their host galaxies. Consequently, we consulted localization studies for these FRBs and subtracted the host galaxy \( DM \) values reported: FRB 190520B \citep{(9)arXiv:2110.07418}, FRB 210117A \citep{(15)arXiv:2211.16790}, FRB 220610A \citep{(23)arXiv:2210.04680}, and FRB 220914A \citep{(22)2023ApJ...949L..26C}. For FRB 220610A, we also adopted the \( DM_{\text{IGM}} \) value provided in its respective study to avoid a negative \( DM_{\text{MW}} \) outcome. Importantly, we note that our main results do not rely on the special treatment of these four FRBs.

To ensure that our main findings are not influenced by the special treatment of these four FRBs, we conducted two additional analyses. First, for all localized FRBs, if a \( DM_{\text{host}} \) value was reported in their localization papers, we adopted that value; otherwise, we used \(30/(1+z) \, \text{pc cm}^{\text{-3}}\). Second, we excluded these four FRBs entirely from the sample. Both results are presented as Alternative 3 and 4 in Table \ref{tab:pearson_results}, demonstrating that our findings remain robust and are not dependent on the special treatment applied to these four unusual FRBs.

Additionally, a negative \(DM_{\text{MW}}\) was obtained for FRB 190611B, suggesting that the FRB's signal may have traversed regions of significantly low density or voids, resulting in a \(DM_{\text{IGM}}\) much lower than the average predicted by the Macquart relation, or that the association between this FRB and its host galaxy may need to be re-examined and redefined, as suggested by \citet{cordes2022redshift}. However, since the presence of this data does not affect our main conclusions, we chose to retain it in our study.
\begin{table}[htbp]\label{tab1}
\centering 
\caption{Localized FRBs Sample} 
\label{tab:localized frbs}       
\renewcommand{\arraystretch}{1.2} 
\begin{adjustwidth}{-\extralength}{0cm}
\begin{tabularx}{\fulllength}{Cccccccccc}

\hline
No. & \multicolumn{1}{c}{FRB} & Telescope & $l$ & $b$ & DM & $z$ & $DM_{\text{MW}}$ \textsuperscript{1} & Repeater? & Ref \\
 & \multicolumn{1}{c}{} &  & (deg) & (deg) & (cm$^{\text{-3}}$ pc) & (redshift) & (cm$^{\text{-3}}$ pc) & (y/n) &  \\
(1) & \multicolumn{1}{c}{(2)} & (3) & (4) & (5) & (6) & (7) & (8) & (9) & (10) \\
\hline
1 & \multicolumn{1}{l|}{FRB121102A} & Arecibo & 174.95 & \text{-0.223} & 570 & 0.1927 & 366.370 & n & \citet{(1)arXiv:1701.01100} \\
2 & \multicolumn{1}{l|}{FRB171020A} & ASKAP & 29.3 & \text{-}51.3 & 114.1 & 0.0087 & 77.196 & n & \citet{(2)arXiv:2305.17960} \\
3 & \multicolumn{1}{l|}{FRB180301A} & Parkes & 204.412 & \text{-}6.481 & 522 & 0.3304 & 208.734 & y & \citet{(3)arXiv:2108.01282} \\
4 & \multicolumn{1}{l|}{FRB180916B} & CHIME & 129.71 & 3.73 & 350.2 & 0.0337 & 291.964 & n & \citet{(4)arXiv:2001.02222} \\
5 & \multicolumn{1}{l|}{FRB180924B} & ASKAP & 0.74247 & \text{-}49.415 & 361.42 & 0.3214 & 56.274 & n & \citet{(5)arXiv:1906.11476} \\
6 & \multicolumn{1}{l|}{FRB181030A} & CHIME & 133.4 & 40.9 & 103.5 & 0.0038 & 70.480 & y & \citet{(6)arXiv:2108.12122} \\
7 & \multicolumn{1}{l|}{FRB181112A} & ASKAP & 342.6 & \text{-}47.7 & 589.27 & 0.4755 & 143.298 & n & \citet{(7)arXiv:1909.11681} \\
8 & \multicolumn{1}{l|}{FRB181220A} & CHIME & 105.24 & \text{-}10.73 & 209.4 & 0.02746 & 157.412 & n & \citet{(8)2024ApJ...971L..51B} \\
9 & \multicolumn{1}{l|}{FRB181223C} & CHIME & 207.75 & 79.51 & 112.51 & 0.03024 & 58.275 & n & \citet{(8)2024ApJ...971L..51B} \\
10 & \multicolumn{1}{l|}{FRB190102C} & ASKAP & 312.65 & \text{-}33.49 & 363.6 & 0.2910 & 85.762 & n & \citet{Macquart2020} \\
11 & \multicolumn{1}{l|}{FRB190418A} & CHIME & 179.3 & \text{-}22.93 & 184.5 & 0.07132 & 96.660 & n & \citet{(8)2024ApJ...971L..51B} \\
12 & \multicolumn{1}{l|}{FRB190520B	\textsuperscript{*}} & FAST & 359.67 & 29.91 & 1204 & 0.2410 & 91.810 & n & \citet{(9)arXiv:2110.07418} \\
13 & \multicolumn{1}{l|}{FRB190523A} & DSA-10 & 117.03 & 44 & 760.8 & 0.6600 & 142.884 & n & \citet{(10)arXiv:1907.01542} \\
14 & \multicolumn{1}{l|}{FRB190608B} & ASKAP & 53.21 & \text{-}48.53 & 338.7 & 0.1178 & 211.982 & n & \citet{Macquart2020} \\
15 & \multicolumn{1}{l|}{FRB190611B} & ASKAP & 312.94 & \text{-}33.28 & 321.4 & 0.3780 & \text{-}35.053 & n & \citet{Macquart2020} \\
16 & \multicolumn{1}{l|}{FRB190711A} & ASKAP & 310.91 & \text{-}33.9 & 593.1 & 0.5220 & 104.044 & y & \citet{Macquart2020} \\
17 & \multicolumn{1}{l|}{FRB190714A} & ASKAP & 289.7 & 49 & 504 & 0.2365 & 274.609 & n & \citet{(11)arXiv:2009.10747} \\
18 & \multicolumn{1}{l|}{FRB191001A} & ASKAP & 341.3 & \text{-}44.8 & 506.92 & 0.2340 & 279.734 & n & \citet{(11)arXiv:2009.10747} \\
19 & \multicolumn{1}{l|}{FRB191228A} & ASKAP & 20.8 & \text{-}64.9 & 297.5 & 0.2430 & 62.368 & n & \citet{(3)arXiv:2108.01282} \\
20 & \multicolumn{1}{l|}{FRB200120E} & CHIME & 142.19 & 41.22 & 87.82 & \text{-}0.0001 & 57.817 & y & \citet{kirsten2022repeating200120} \\
21 & \multicolumn{1}{l|}{FRB200430A} & ASKAP & 17.06 & 52.52 & 380.1 & 0.1600 & 217.390 & n & \citet{(11)arXiv:2009.10747} \\
22 & \multicolumn{1}{l|}{FRB200906A} & ASKAP & 202.4 & \text{-}49.77 & 577.8 & 0.3688 & 229.724 & n & \citet{(3)arXiv:2108.01282} \\
23 & \multicolumn{1}{l|}{FRB201123A} & MeerKAT & 340.23 & \text{-}9.68 & 433.9 & 0.0507 & 363.022 & n & \citet{(13)arXiv:2205.14600} \\
24 & \multicolumn{1}{l|}{FRB201124A} & CHIME & 177.6 & \text{-}8.5 & 410.83 & 0.0979 & 300.862 & y & \citet{(14)arXiv:2106.11993} \\
25 & \multicolumn{1}{l|}{FRB210117A \textsuperscript{*}} & ASKAP & 45.79 & \text{-}57.59 & 730 & 0.2140 & 54.991 & n & \citet{(15)arXiv:2211.16790} \\
26 & \multicolumn{1}{l|}{FRB210320C} & ASKAP & 318.88 & 45.31 & 384.593 & 0.2797 & 116.857 & n & \citet{(16)arXiv:2302.05465} \\
27 & \multicolumn{1}{l|}{FRB210405I} & MeerKAT & 338.19 & \text{-}4.59 & 566.43 & 0.0660 & 482.984 & n & \citet{(17)arXiv:2302.09787} \\
28 & \multicolumn{1}{l|}{FRB210410D} & MeerKAT & 312.32 & \text{-}34.13 & 578.78 & 0.1415 & 431.925 & n & \citet{(18)arXiv:2302.09754} \\
29 & \multicolumn{1}{l|}{FRB210807D} & ASKAP & 39.81 & \text{-}14.89 & 251.9 & 0.1293 & 115.436 & n & \citet{(16)arXiv:2302.05465} \\
30 & \multicolumn{1}{l|}{FRB211127I} & ASKAP & 311.99 & 43.56 & 234.83 & 0.0469 & 167.058 & n & \citet{(19)arXiv:2305.14863} \\
31 & \multicolumn{1}{l|}{FRB211203C} & ASKAP & 314.43 & 30.47 & 636.2 & 0.3439 & 310.726 & n & \citet{(16)arXiv:2302.05465} \\
32 & \multicolumn{1}{l|}{FRB211212A} & ASKAP & 243.95 & 47.76 & 206 & 0.0707 & 118.672 & n & \citet{(16)arXiv:2302.05465} \\
33 & \multicolumn{1}{l|}{FRB220105A} & ASKAP & 18.84 & 74.68 & 583 & 0.2785 & 316.335 & n & \citet{(16)arXiv:2302.05465} \\
34 & \multicolumn{1}{l|}{FRB220207C} & DSA-110 & 106.94 & 18.39 & 263 & 0.0430 & 198.376 & n & \citet{(20)arXiv:2307.03344} \\
35 & \multicolumn{1}{l|}{FRB220307B} & DSA-110 & 116.24 & 10.47 & 499.328 & 0.2481 & 259.665 & n & \citet{(20)arXiv:2307.03344} \\
36 & \multicolumn{1}{l|}{FRB220310F} & DSA-110 & 140.02 & 34.8 & 462.657 & 0.4780 & 14.413 & n & \citet{(20)arXiv:2307.03344} \\
37 & \multicolumn{1}{l|}{FRB220319D} & DSA-110 & 129.18 & 9.11 & 110.95 & 0.0110 & 72.187 & n & \citet{(21)arXiv:2301.01000} \\
38 & \multicolumn{1}{l|}{FRB220418A} & DSA-110 & 110.75 & 44.47 & 624.124 & 0.6220 & 41.809 & n & \citet{(20)arXiv:2307.03344} \\
39 & \multicolumn{1}{l|}{FRB220506D} & DSA-110 & 108.35 & 16.51 & 396.651 & 0.3004 & 110.398 & n & \citet{(20)arXiv:2307.03344} \\
40 & \multicolumn{1}{l|}{FRB220509G} & DSA-110 & 100.94 & 25.48 & 270.26 & 0.0894 & 167.399 & n & \citet{(22)2023ApJ...949L..26C} \\
41 & \multicolumn{1}{l|}{FRB220610A \textsuperscript{*}} & ASKAP & 8.87 & \text{-}70.13 & 1458.1 & 1.0160 & 6.275 & n & \citet{(23)arXiv:2210.04680} \\
42 & \multicolumn{1}{l|}{FRB220825A} & DSA-110 & 106.99 & 17.79 & 649.893 & 0.2414 & 416.178 & n & \citet{(20)arXiv:2307.03344} \\
43 & \multicolumn{1}{l|}{FRB220912A} & CHIME & 347.27 & 48.7 & 221.8 & 0.0771 & 126.832 & y & \citet{(24)arXiv:2211.09049} \\
44 & \multicolumn{1}{l|}{FRB220914A \textsuperscript{*}} & DSA-110 & 104.31 & 26.13 & 630.703 & 0.1125 & 159.697 & n & \citet{(22)2023ApJ...949L..26C} \\
45 & \multicolumn{1}{l|}{FRB220920A} & DSA-110 & 104.92 & 38.89 & 314.98 & 0.1582 & 153.784 & n & \citet{(20)arXiv:2307.03344} \\
46 & \multicolumn{1}{l|}{FRB221012A} & DSA-110 & 101.14 & 26.14 & 440.36 & 0.2847 & 168.185 & n & \citet{(20)arXiv:2307.03344} \\
47 & \multicolumn{1}{l|}{FRB230718A} & ASKAP & 259.66 & \text{-}1.03 & 477 & 0.0357 & 418.343 & n & \citet{(25)arXiv:2311.16808} \\
\hline
\end{tabularx}
\begin{flushleft}
\footnotesize{Notes: \textsuperscript{1}Values derived in this study. 

\textsuperscript{*}\( DM_{\text{MW}} \) values for these four FRBs are obtained from works dedicated to their localization.}
\end{flushleft}
\end{adjustwidth} \end{table}

\section{Analysis and Findings} \label{sec:3}

Currently, X-ray observations provide one of the most effective methods to study the hot gas distribution in and around the Milky Way, particularly through the detection of O \textsc{vii} and O \textsc{viii} absorption and emission lines, which are sensitive to the temperature and density of gas in the Galactic disk and halo. In this section, we compare our $DM_{\text{MW}}$ sample with these X-ray tracers to investigate potential correlations and the role they play in mapping the hot gas content of the Milky Way.

\subsection{\texorpdfstring{O \textsc{vii} Absorption}{O VII Absorption}}
In this subsection, we analyze the relationship between \(DM_{\text{MW}}\) and O \textsc{vii} absorption lines, which trace hot gas in both the Galactic disk and halo, using active galactic nuclei (AGN) as background sources. We used the dataset from \citet{fang2015xmm} but excluded the source Mkn 841, as its O \textsc{vii} absorption line equivalent width significantly exceeds typical Galactic levels and is likely contaminated, as suggested by \citet{longinotti2010high}.

\begin{figure}[H]
\begin{adjustwidth}{-\extralength}{0cm}
    \centering
    \begin{minipage}{0.65\textwidth}  
        \centering
        \includegraphics[width=\linewidth]{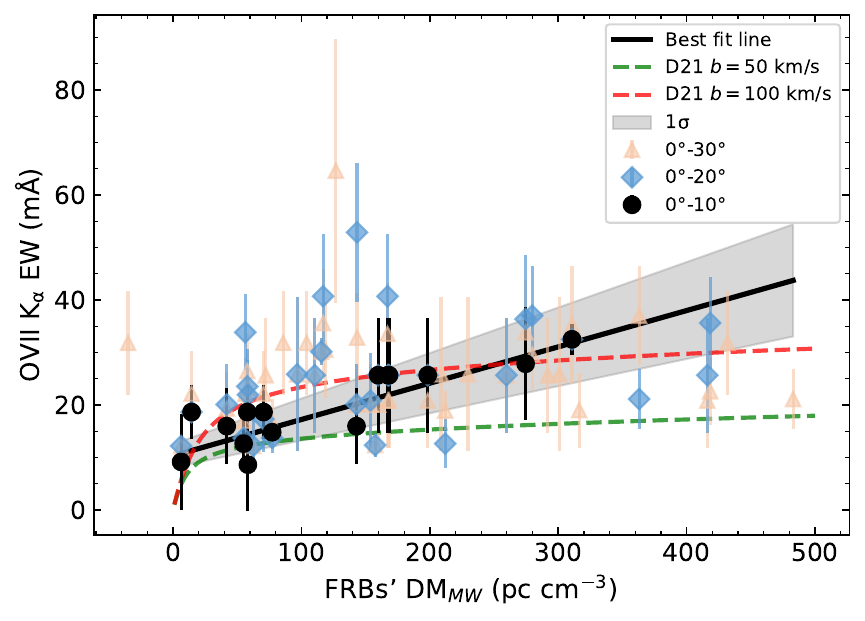}
    \end{minipage}
    \hfill
    \begin{minipage}{0.65\textwidth}  
        \centering
        \includegraphics[width=\linewidth]{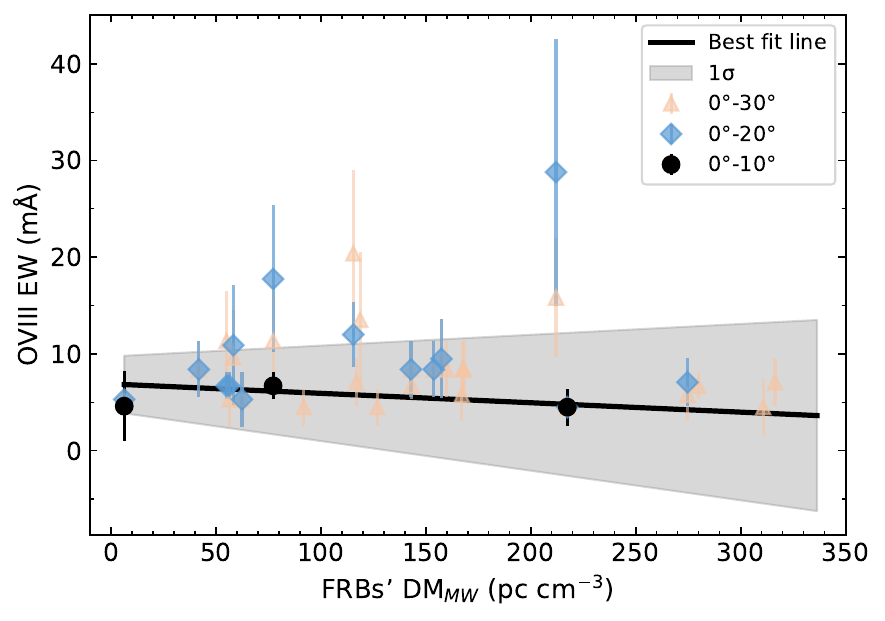}
    \end{minipage}
\end{adjustwidth}
    \caption{The correlation between \( DM_{\text{MW}} \) and the equivalent width (EW) of O \textsc{vii} (left) and O \textsc{viii} (right) absorption lines. For each FRB, black circles, blue squares, and light orange triangles represent the average equivalent width of sources found within 10, 20, and 30-degree angular regions centered on the FRB's Galactic coordinates (\( l, b \)). The solid lines show the best-fit results obtained from the 10-degree region data for both O \textsc{vii} and O \textsc{viii} using MCMC fitting, with shaded regions representing the 1\(\sigma\) uncertainties. In the left panel, the green and red dashed lines correspond to Doppler parameters \( b = 50 \, \text{km/s} \) and \( 100 \, \text{km/s} \), respectively, based on the empirical formula derived by \citet{Das2021empirical} (see Section \ref{sec:4.1} for details).
 \label{fig:absorption}}
\end{figure}

The left panel of Figure \ref{fig:absorption} shows the correlation between the equivalent width of O \textsc{vii} absorption lines and \(DM_{\text{MW}}\). For each FRB, we identified O \textsc{vii} absorption sources within 10, 20, and 30-degree angular regions centered on their respective sky positions in Galactic coordinates, represented by black circles, blue squares, and light orange triangles, respectively. For each angular region of each FRB, the equivalent widths of the identified sources were averaged to obtain a representative value for that region. A total of 15 FRBs had O \textsc{vii} absorption sources located within the 10-degree region. A Pearson correlation analysis of 15 data points between O \textsc{vii} equivalent width (EW) and \(DM_{\text{MW}}\) yielded an \(r\)-value of 0.8636 and a \(p\)-value of \(3.35 \times 10^{\text{-5}}\), indicating a strong linear relationship. Using this correlation, we applied Markov Chain Monte Carlo (MCMC) fitting to derive an empirical relation for estimating \(DM_{\text{MW}}\) based on the equivalent width of the O \textsc{vii} absorption line:

\begin{equation}
DM_{\text{MW}} = (288.71 \pm 73.93) \left( \dfrac{EW_{\text{O\,\textsc{vii}}}}{20\, \text{m\AA}} \right) \text{-} (141.17 \pm 50.34)\, \text{pc\,cm}^{\text{-3}}.
\label{eq:dm_mw_ovii}
\end{equation}

In addition to this analysis, we also examined correlations between the equivalent width of O \textsc{viii} absorption lines (\(EW_{\text{O\,VIII}}\)), and the emission intensities of O \textsc{vii} (\(I_{\text{O\,VII}}\)) and O \textsc{viii} (\(I_{\text{O\,VIII}}\)), with \(DM_{\text{MW}}\). The equivalent width (\(EW\)) reflects the strength of absorption lines, while the emission intensity (\(I\)) quantifies the brightness of the corresponding emission lines. The correlations were analyzed across three angular regions for the absorption lines (0°–10°, 0°–20°, and 0°–30°) and for the emission intensities (0°–5° and 0°–10°). The results are summarized in Table \ref{tab:pearson_results}.

\begin{table}[htbp]
\caption{Pearson Correlation Analysis between \(DM_{\text{MW}}\) and Absorption/Emission Line Properties (O \textsc{vii}/O \textsc{viii} Equivalent Width and Emission Intensity) across Different Angular Regions in Galactic Coordinates. Results for alternative analyses to test the robustness of the correlation are also included.}
\label{tab:pearson_results}
\renewcommand{\arraystretch}{1.2}
\begin{adjustwidth}{-\extralength}{0cm}
\begin{tabularx}{0.8\fulllength}{cccccc}
\hline
\textbf{Analysis} & \textbf{Angular Region} & \textbf{\(DM_{\text{IGM}}\)} & \textbf{\(DM_{\text{host}}\)} & \textbf{\(R\)-value} & \textbf{\(P\)-value} \\
                  & (degree)                &                              & (\(\text{pc\,cm}^{\text{-3}}\))     &                       &                       \\ 
\hline
\multirow{3}{*}{\(EW_{\text{O\,VII}}\)} 
  & 0°–10°  & Macquart-relation & \(30/(1+z)\) & 0.8636 & \(3.35 \times 10^{\text{-5}}\) \\
  & 0°–20°  & Macquart-relation & \(30/(1+z)\) & 0.3436 & 0.0584 \\
  & 0°–30°  & Macquart-relation & \(30/(1+z)\) & 0.1029 & 0.5061 \\
\hline
\multirow{3}{*}{\(EW_{\text{O\,VIII}}\)} 
  & 0°–10°  & Macquart-relation & \(30/(1+z)\) & \text{-}0.2255 & 0.8552 \\
  & 0°–20°  & Macquart-relation & \(30/(1+z)\) & 0.2422  & 0.4041 \\
  & 0°–30°  & Macquart-relation & \(30/(1+z)\) & \text{-}0.1328 & 0.5178 \\
\hline
\multirow{2}{*}{\(I_{\text{O\,VII}}\)}   
  & 0°–5°   & Macquart-relation & \(30/(1+z)\) & 0.4849  & \(9.84 \times 10^{\text{-4}}\) \\
  & 0°–10°  & Macquart-relation & \(30/(1+z)\) & 0.5396  & \(9.06 \times 10^{\text{-5}}\) \\
\hline
\multirow{2}{*}{\(I_{\text{O\,VIII}}\)}  
  & 0°–5°   & Macquart-relation & \(30/(1+z)\) & 0.5319  & \(4.92 \times 10^{\text{-4}}\) \\
  & 0°–10°  & Macquart-relation & \(30/(1+z)\) & 0.4546  & 0.0013 \\
\hline
\textsuperscript{*}Alternative 1 & 0°–10° & Macquart-relation & \(60/(1+z)\) & 0.8512 & \(5.71 \times 10^{\text{-5}}\) \\
\textsuperscript{*}Alternative 2 & 0°–10° & Macquart-relation & \(100/(1+z)\) & 0.8144 & \(2.19 \times 10^{\text{-4}}\) \\
\textsuperscript{*}Alternative 3 & 0°–10° & Macquart-relation & Localization papers\textsuperscript{1} & 0.8122 & \(2.35 \times 10^{\text{-4}}\) \\
\textsuperscript{*}Alternative 4\textsuperscript{2} & 0°–10° & Macquart-relation & \(30/(1+z)\) & 0.8395 & \(6.36 \times 10^{\text{-4}}\) \\
\textsuperscript{*}Alternative 5 & 0°–10° & \citet{zhang2018fast} & \(30/(1+z)\) & 0.8630 & \(3.43 \times 10^{\text{-5}}\) \\
\hline
\end{tabularx}
\end{adjustwidth}

\footnotesize{Notes: \textsuperscript{*} All alternative analyses were based on the \(DM_{\text{MW}}\) and \(EW_{\text{O\,VII}}\) within the 10-degree regions. \\
\textsuperscript{1} For FRBs whose localization papers do not provide \(DM_{\text{host}}\), we adopt \(30/(1+z)\). \\
\textsuperscript{2} This analysis excludes 4 unusual FRBs.}
\end{table}

\subsection{\texorpdfstring{O \textsc{viii} Absorption}{O VIII Absorption}}
Next, we investigated the relationship between \(DM_{\text{MW}}\) and O \textsc{viii} absorption lines. Similar to O \textsc{vii}, O \textsc{viii} absorption lines can trace hot gas in both the Galactic disk and halo. We compiled all available O \textsc{viii} absorption data that use AGN as background sources \citep{Zappacosta2010ApJ...717...74Z,Gupta_2012,FangandJiang2014ApJ...785L..24F,Bonamente2016MNRAS.457.4236B,Das2019ApJ...882L..23D}.

The right panel of Figure \ref{fig:absorption} shows the correlation between the equivalent width of O \textsc{viii} absorption lines and \(DM_{\text{MW}}\). Similar to the left panel, for each FRB, we searched for O \textsc{viii} absorption sources within 10, 20, and 30-degree angular regions. However, due to the limited O \textsc{viii} data, only three FRBs have O \textsc{viii} absorption data points within the 10-degree region. Even within the 30-degree region, some FRBs lack O \textsc{viii} absorption sources. Although we performed MCMC fitting and Pearson correlation analysis using data from the 10-degree region, which yielded an \(r\)-value of \text{-}0.2255 and a \(p\)-value of  0.8552, these results are not statistically significant.

\subsection{\texorpdfstring{O \textsc{vii} and O \textsc{viii} Emission}{OVII and O VIII Emission}}

In addition to absorption line data, O \textsc{vii} and O \textsc{viii} emission line data provide crucial information about the distribution of hot gas in the Milky Way. In this subsection, we discuss the emission line data and their correlation with \(DM_{\text{MW}}\), using the most recent dataset published by \citet{quovii2024}.

\begin{figure}[ht!]
\begin{adjustwidth}{-\extralength}{0cm}
    \centering
    \begin{minipage}{0.65\textwidth}  
        \centering
        \includegraphics[width=\linewidth]{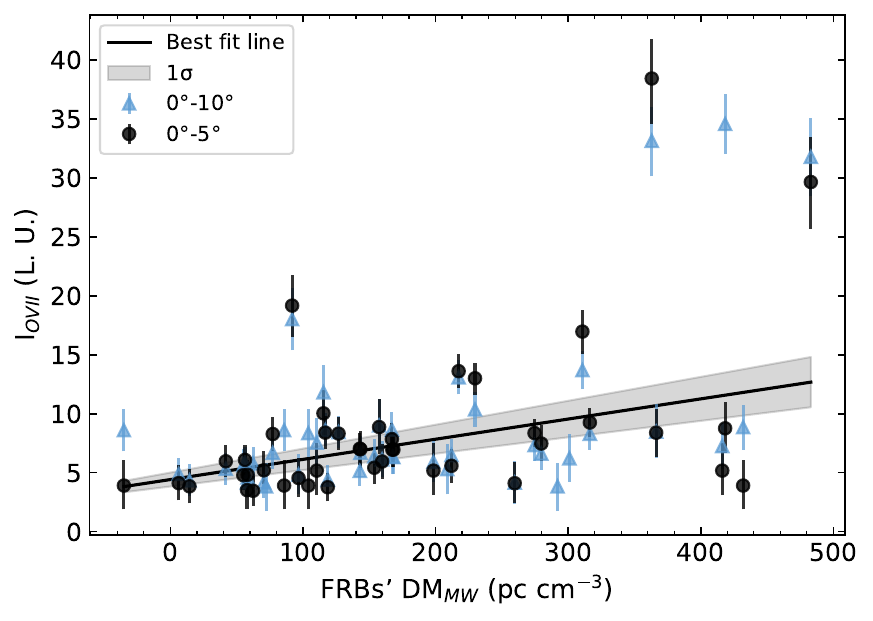}  
    \end{minipage}
    \hfill
    \begin{minipage}{0.65\textwidth}  
        \centering
        \includegraphics[width=\linewidth]{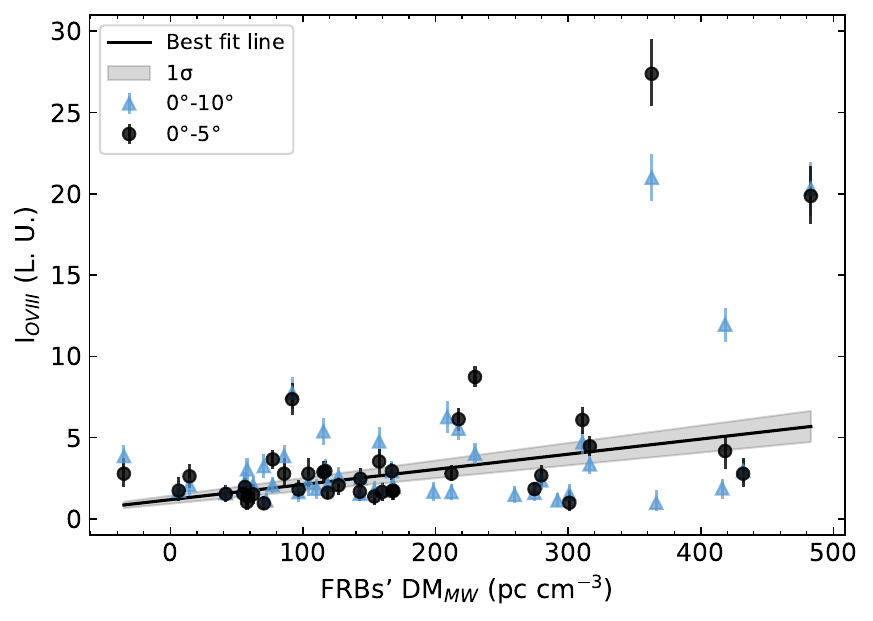}  
    \end{minipage}
\end{adjustwidth}
    \caption{The correlation between \(DM_{\text{MW}}\) and the emission intensities of O \textsc{vii} (left) and O \textsc{viii} (right) measured in Line Units (L.U.). For each FRB, black circles and blue triangles represent the average emission intensities within 5-degree and 10-degree angular regions centered on the FRB's Galactic coordinates (\(l, b\)), respectively. The solid lines show the best-fit results obtained from the 5-degree region data for both O \textsc{vii} and O \textsc{viii} using MCMC fitting. The shaded regions represent the 1$\sigma$ uncertainty associated with these fits. 
    \label{fig:emission}}
\end{figure}

Figure \ref{fig:emission} illustrates the correlation between the emission intensities of O \textsc{vii} and O \textsc{viii} (measured in Line Units, L.U.) and \(DM_{\text{MW}}\). We averaged the emission data points within 5-degree and 10-degree regions centered on the Galactic coordinates of each FRB. The left panel shows the correlation between O \textsc{vii} emission and \(DM_{\text{MW}}\), while the right panel shows the same for O \textsc{viii}. Black circles represent the average values of data points within the 5-degree region, and blue triangles represent the 10-degree averages. For O \textsc{vii} emission, Pearson correlation analysis on 43 data points within the 5-degree region yielded an \(r\)-value of 0.4849 and a \(p\)-value of \(9.84 \times 10^{\text{-4}}\). For O \textsc{viii}, analysis of 39 data points in the 5-degree region gave an \(r\)-value of 0.5319 and a \(p\)-value of \(4.92 \times 10^{\text{-4}}\). Although both correlations exhibit relatively weak relationships, these correlations are primarily driven by a few outliers with larger values. Therefore, we consider these correlations to be unreliable.

\section{Discussion} \label{sec:4}

\subsection{\texorpdfstring{Why O \textsc{vii} Absorption Shows the Strongest Correlation}{Why O VII Absorption Shows the Strongest Correlation}}\label{sec:4.1}
Our analysis reveals that the strongest correlation with \(DM_{\text{MW}}\) is observed for the O \textsc{vii} absorption line data, as shown in Table \ref{tab:pearson_results}. This can be attributed to the environment of the Galactic halo, where temperatures around \(2 \times 10^6 \, \text{K}\) are ideal for the presence of highly ionized metals \citep{Kaaret2020NatAs...4.1072K}. In this temperature range, O \textsc{vii} becomes the dominant ion, as evidenced by absorption lines detected in the soft X-ray band by observatories such as \textit{Chandra} and \textit{XMM-Newton} \citep{nicastro2005Natur.433..495N,Bregman2007ApJ...669..990B}. Naturally, O \textsc{vii} absorption traces both the hot gas in the Galactic halo and the Galactic disk, reflecting contributions from both components. The high ionization fraction of O \textsc{vii} at these temperatures makes it an effective tracer of the overall hot gas in the Milky Way.

Recent work by \citet[][D21]{Das2021empirical} divided the total \( DM_{\text{MW}} \) contribution from the Milky Way into four components, corresponding to different gas temperature ranges: cold, cool, warm, and hot phases. Their analysis found that the majority of the \( DM_{\text{MW}} \) contribution originates from the hot gas phase. Similar to previous studies \citep[e.g.,][]{Shull_2018,prochaska2019probing,yamasaki2020galactic}, they used O \textsc{vii} absorption column density to estimate \( DM_{\text{MW}} \). By incorporating key parameters, such as Galactic metallicity and solar oxygen abundance, they developed an empirical formula linking \( DM_{\text{MW}} \) to the observed O \textsc{vii} absorption.

Following their approach, we used the same formula to convert the relationship into \( DM \) versus equivalent width (EW) by assuming Doppler parameters of \( b = 50 \, \text{km/s} \) and \( 100 \, \text{km/s} \), which are represented in the left panel of Figure \ref{fig:absorption} by the green and red dashed lines, respectively.

This method aligns closely with our findings, as all three studies emphasize the utility of O \textsc{vii} absorption as a reliable tracer for estimating the Milky Way's \( DM \) contribution. Notably, our \( DM_{\text{MW}} \) data points, also shown in Figure \ref{fig:absorption}, are almost entirely contained within the range defined by these two lines, further supporting the robustness of this approach. These results underscore the critical role of O \textsc{vii} absorption data in tracing the hot gas distribution and quantifying its contribution to the overall \( DM \).

In contrast, the weaker correlation between O \textsc{viii} and \(DM_{\text{MW}}\) can be explained by its higher ionization temperature requirement (approximately \(10^{6.5}\) K), which is less common in the Galactic halo compared to the 1–2 million K temperature range optimal for O \textsc{vii}. This scarcity of high-temperature gas limits the available data points for O \textsc{viii}, as evidenced by the small sample of only three FRBs within a 5-degree region. Additionally, O \textsc{viii} column densities are highly sensitive to the Doppler-b parameter, which is poorly constrained by current observations. In contrast, O \textsc{vii}, which forms more readily under typical halo conditions, serves as a more reliable tracer of the hot gas \citep[see, e.g.,][]{Mathur2003ApJ...582...82M,Williams2005ApJ...631..856W}.

Regarding the O \textsc{vii} and O \textsc{viii} emission lines, although Pearson analysis indicates positive correlations with \(DM_{\text{MW}}\), these correlations are weaker compared to those observed for O \textsc{vii} absorption. This discrepancy likely arises because emission lines primarily trace regions of higher density, such as the Galactic disk, which may explain the weaker correlation with the dispersion measure \citep{henley2012xmm}.

However, given the small number of localized FRBs, it is important to acknowledge that the limited size of our \(DM_{\text{MW}}\) sample within 10-degree regions of O \textsc{vii} absorption contributes to the uncertainty in the observed positive correlation. We emphasize that our results merely suggest a potential connection between \(DM_{\text{MW}}\) and the equivalent width of O \textsc{vii} absorption. Future work, such as that conducted by CHIME/FRB \citep{amiri2018chime}, is expected to significantly expand the sample of localized FRBs, facilitating a more reliable and comprehensive analysis of this correlation.

\subsection{Comparison with MW Electron Density Distribution Models}

In this subsection, we compare the empirically derived \(DM_{\text{MW}}\) values from our localized FRB sample with \(DM_{\text{model}}\) predictions from several Milky Way electron density distribution models. These models estimate the Milky Way’s contribution to the total dispersion measure based on varying assumptions about the electron distribution. Specifically, we examine the models proposed by \citet[][F13]{fang2013ApJ...762...20F}, \citet[][T17]{Troitsky2017MNRAS.468L..36T}, and \citet[][M22]{Martynenko2022MNRAS.511..843M}, along with the NE2001 and YMW16 disk models, which provide lower-limit estimates.

Since T17 and M22 primarily focus on describing the electron density distribution in the Galactic halo and do not provide a corresponding disk model, we adopt a disk electron density model that follows a plane-parallel distribution:

\begin{equation}
    n_{e}(z) = n_{0}e^{-\left | z \right |/z_{0}}   
\end{equation}

where \(n_0\) and \(z_0\) represent the mid-plane disk electron density and the scale height, respectively. Based on the analysis of 37 pulsar sightlines, \citet[][O20]{Ocker2020ApJ...897..124O} determined best-fit parameters of \(n_0 = 0.015 \pm 0.001 \, \text{cm}^{\text{-3}}\) and \(z_0 = 1.57^{\text{+0.15}}_{\text{-0.14}} \, \text{kpc}\), which we adopt here.

\begin{figure}[H]
\centering
\includegraphics[width=10.5cm]{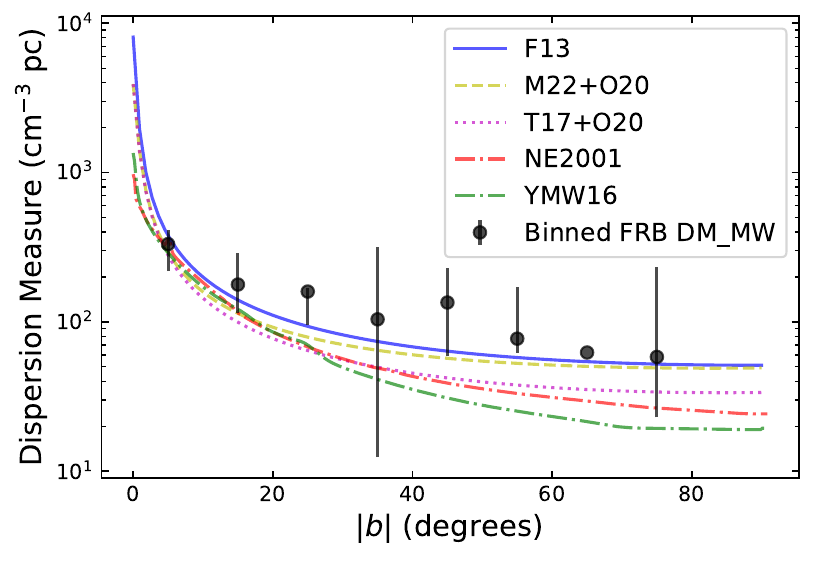}
\caption{Relationship between DM and \(|b|\). The lines represent the three Milky Way electron density distribution models. The solid blue line represents the model by F13, while the pink dotted and yellow dashed lines represent the models by T17 and M22, respectively, both combined with the plane-parallel disk component from O20. The red and green dashdot lines depict two widely used disk electron density models, NE2001 and YMW16, serving as lower limits. Black points indicate the median values of our sample across bins.}
\label{fig:model-b}
\end{figure}

Figure \ref{fig:model-b} illustrates the relationship between \( DM \) and Galactic latitude \(|b|\). The lines in the figure represent various models of the Milky Way's electron density distribution. YMW16 and NE2001, as the most widely used disk electron density models, are also shown as lower limits. 

The model values were calculated at 30-degree intervals in longitude \(|l|\), following a longitude range similar to that used by \citet{nakashima2018spatial}, and then averaged, with the curves representing these average values. The \( DM_{\text{MW}} \) sample was divided into 9 bins, each spanning 10 degrees in \(|b|\), and is shown as black points. Each point represents the median \( DM_{\text{MW}} \) value within the corresponding bin, with error bars indicating the 16th and 84th percentiles.

Overall, our \(DM_{\text{MW}}\) values are consistent with the general trend of the models, showing a decrease with increasing \(|b|\). Moreover, for disk and halo combination models, such as F13 and M22+O20, our results demonstrate agreement within the uncertainties. In contrast, for pure disk-only models like NE2001 and YMW16, our \(DM_{\text{MW}}\) sample exhibits significant excess, emphasizing the contribution of the halo component. We noted that for \(|b|\) values between 20 and 50 degrees, \(DM_{\text{MW}}\) shows a noticeable excess compared to the models, suggesting that the models may not fully capture the complexity of the Milky Way's electron density distribution or that additional factors influence the observed values.

One likely factor is the simplified assumption for the host galaxy contribution, where a constant value of \( 30/(1+z) \, \text{pc cm}^{\text{-3}} \) was used. Such an assumption does not account for the expected variability in the host galaxy’s contribution\citep{mo2023MNRAS.518..539M}. Additionally, the sparsity of data may also be a contributing factor, for instance, in the \(|b|\) range of 50–60 degrees, there are only three data points, which also explains the asymmetry in the error bars. 

Furthermore, uncertainties in the Macquart relation used to estimate \( DM_{\text{IGM}} \) may also contribute to the observed excess. These uncertainties arise from the scatter in the baryon fraction and the distribution of free electrons in the intergalactic medium. Together, these factors could lead to systematic biases in the derived \( DM_{\text{MW}} \) values.

To address this, we also tested an alternative \( DM_{\text{IGM}}-z \) relation based on the model from \citet{deng2014cosmological}, using parameters from \citet{zhang2018fast}. This approach yielded a slightly smaller slope for Equation \ref{eq:dm_mw_ovii}, which could slightly reduce the excess \( DM_{\text{MW}} \) in comparison to Milky Way electron density models, as presented as Alternative 5 in Table \ref{tab:pearson_results}. Moreover, we reached the same conclusion that there is a positive correlation between O \textsc{vii} absorption and \( DM_{\text{MW}} \).

\section{Summary} \label{sec:summary}

In this paper, we investigated the total \(DM_{\text{MW}}\), which reflects the contributions of the Milky Way’s disk and halo to its electron density, using O \textsc{vii} and O \textsc{viii} absorption and emission data as tracers of the hot, highly ionized gas. Our analysis indicates a positive correlation between O \textsc{vii} absorption and \(DM_{\text{MW}}\), from which we derive an empirical relation (Equation~\ref{eq:dm_mw_ovii}) to estimate \(DM_{\text{MW}}\) from the equivalent width of O \textsc{vii} absorption. This relation highlights O \textsc{vii} absorption as a robust tracer of the Milky Way’s hot gas and offers a practical tool for future FRB-based studies to constrain the Milky Way’s DM contribution.

Our findings also indicate that \(DM_{\text{MW}}\) reflects contributions from both the disk and halo components of the Milky Way. This conclusion is supported by the weaker correlation between \(DM_{\text{MW}}\) and O \textsc{vii}/O \textsc{viii} emission, which primarily traces dense disk regions, and by comparisons with the MW electron density models. The latter shows that \(DM_{\text{MW}}\) aligns better with models incorporating both disk and halo components, such as F13, M22+O20, but significantly exceeds predictions from pure disk models such as NE2001 and YMW16, emphasizing the importance of the halo’s contribution. 

Furthermore, the lack of a strong correlation with O \textsc{viii} absorption suggests that the primary temperature of the Milky Way’s hot gas is predominantly around \(2 \times 10^6\) K, consistent with O \textsc{vii} absorption. While gas at higher temperatures (\(3\)--\(5 \times 10^6\) K), typically associated with O \textsc{viii} absorption, is present, it appears to be less abundant.

In summary, our study highlights the potential of O \textsc{vii} absorption data as a reliable tracer of the Milky Way’s hot gas, providing new insights into \( DM_{\text{MW}} \) contributions. While our results are broadly consistent with previous findings, uncertainties remain, particularly regarding spatial variations in absorption and emission data across different sky regions. 

Future studies on FRBs can leverage our empirical formula to estimate \( DM_{\text{MW}} \) if O \textsc{vii} absorption sources are detected in the vicinity of an FRB. This approach offers a practical tool for the initial localization of FRBs and serves as a stepping stone for further studies on missing baryons. These efforts will not only refine our understanding of Galactic and extragalactic environments but also advance the broader investigation of FRB sources and the cosmic baryon distribution.

\section*{}
\noindent
\textbf{Author Contributions:} Data curation, Writing—original draft preparation, Formal analysis, J.W.; Methodology, Z.Z.; Software, X.J.; Conceptualization, Methodology, Supervision, Writing—review and editing, T.F. All authors have read and agreed to the published version of the manuscript.

\vspace{0.5\baselineskip} 

\noindent
\textbf{Conflict of Interest:} The authors declare no conflicts of interest.

\vspace{0.5\baselineskip} 

\noindent
\textbf{Data Availability Statement:} This study does not include new data. All data used are from previously published sources, as cited in the manuscript.

\vspace{0.5\baselineskip} 

\noindent
\textbf{Funding:} J.W., Z.Z., and T.F. were supported by the National Natural Science Foundation of China under Nos. 11890692, 12133008, 12221003. We also acknowledge the science research grant from the China Manned Space Project with No. CMS-CSST-2021-A04.

\begin{adjustwidth}{-\extralength}{0cm}

\reftitle{References}

\PublishersNote{}
\end{adjustwidth}
\end{document}